\documentclass[mathleft
]{an}
\usepackage{graphicx}
\usepackage{times}
\usepackage{url,hyperref,breakurl}	
\begin{document}

\Pagespan{345}{}
\Yearpublication{2014}%
\Yearsubmission{2014}%
\Month{04}%
\Volume{335}%
\Issue{4}%

\title{Investigation of a transiting planet candidate in Trumpler 37: an astrophysical false positive eclipsing spectroscopic binary star\thanks{Based in part on data collected at Subaru Telescope, which is operated by the National Astronomical Observatory of Japan. 
 Some of the data presented herein were obtained at the W.M. Keck Observatory, which is operated as a scientific partnership among the California Institute of Technology, the University of California and the National Aeronautics and Space Administration (Proposal ID H215Hr). 
 The Observatory was made possible by the generous financial support of the W.M. Keck Foundation. 
 Based on observations obtained with telescopes of the University Observatory Jena, which is operated by the Astrophysical Institute of the Friedrich-Schiller-University.
 Based on observations collected at the Centro Astron\'{o}mico Hispano Alem\'{a}n (CAHA) at Calar Alto, operated jointly by the Max-Planck Institut f\"ur Astronomie and the Instituto de Astrof\'{i}sica de Andaluc\'{i}a (CSIC, Proposal IDs H10-3.5-015 and H10-2.2-004).
 Some of the observations reported here were obtained at the MMT Observatory, a joint facility of the Smithsonian Institution and the University of Arizona (Proposal ID 2010c-SAO-5).
}}

\author{R. Errmann\inst{1,2}\fnmsep\thanks{Corresponding author:
  \email{ronny.errmann@uni-jena.de}\newline}
\and G. Torres \inst{3}
\and T.O.B. Schmidt \inst{1}
\and M. Seeliger \inst{1}
\and A.W. Howard \inst{4}
\and G. Maciejewski \inst{5}
\and R. Neuh\"auser \inst{1}
\and S. Meibom \inst{3}
\and A. Kellerer\inst{6,7}
\and D.P. Dimitrov\inst{8}
\and B. Dincel \inst{1}
\and C. Marka \inst{1}
\and M. Mugrauer \inst{1}
\and Ch. Ginski \inst{1}
\and Ch. Adam \inst{1}
\and St. Raetz \inst{1}
\and J.G. Schmidt \inst{1}
\and M.M. Hohle \inst{1}
\and A. Berndt \inst{1}
\and M. Kitze \inst{1}
\and L. Trepl \inst{1}
\and M. Moualla \inst{1}
\and T. Eisenbei{\ss} \inst{1}
\and S. Fiedler \inst{1}
\and A. Dathe \inst{1}
\and Ch. Graefe \inst{9}
\and N. Pawellek \inst{1}
\and K. Schreyer \inst{1}
\and D.P. Kjurkchieva\inst{10}
\and V.S. Radeva\inst{10}
\and V. Yotov\inst{10}
\and W.P. Chen \inst{11}
\and S.C.-L. Hu \inst{11,12}
\and Z.-Y. Wu \inst{13}
\and X. Zhou \inst{13}
\and T. Pribulla \inst{14}
\and J. Budaj \inst{14,15}
\and M. Va\v{n}ko \inst{14}
\and E. Kundra \inst{14}
\and \v{L}. Hamb\'alek \inst{14}
\and V. Krushevska\inst{16}
\and {\L}. Bukowiecki \inst{5}
\and G. Nowak \inst{17,18}
\and L. Marschall\inst{19}
\and H. Terada \inst{20}
\and D. Tomono \inst{20}
\and M. Fernandez \inst{21}
\and A. Sota \inst{21}
\and H. Takahashi\inst{22}
\and Y. Oasa\inst{23}
\and C. Brice\~no\inst{24,25}
\and R. Chini\inst{26,27}
\and C.H. Broeg \inst{28}
}
\titlerunning{Transiting candidate in Trumpler\,37}
\authorrunning{Errmann et al.}
\institute{
Astrophysikalisches Institut und Universit\"ats-Sternwarte, Schillerg\"a{\ss}chen 2-3, 07745 Jena, Germany
\and 
Abbe Center of Photonics, Friedrich-Schiller-Universit\"at Jena, Max-Wien-Platz 1, 07743 Jena, Germany
\and 
Harvard-Smithsonian Center for Astrophysics, 60 Garden St., Mail Stop 20, Cambridge MA 02138, USA
\and 
Institute for Astronomy, University of Hawaii - Manoa, 2680 Woodlawn Dr., Honolulu, HI 96822, USA
\and 
Centre for Astronomy, Faculty of Physics, Astronomy and Informatics, Nicolaus Copernicus University, Grudziadzka 5, 87-100 Torun, Poland
\and 
Department of Physics, Durham University, South Road, Durham DH1 3LE, United Kingdom
\and 
Institute of Astronomy, University of Hawaii, 640 N. A'ohoku Place, Hilo, Hawaii 96720
\and 
Institute of Astronomy and NAO, Bulg. Acad. Sci., 72 Tsarigradsko Chaussee Blvd., 1784 Sofia, Bulgaria
\and 
Christian-Albrechts-Universit\"at Kiel, Leibnizstra\ss e 15, D-24098 Kiel, Germany
\and 
Shumen University, 115 Universitetska str., 9700 Shumen, Bulgaria
\and 
Graduate Institute of Astronomy, National Central University, Jhongli 32001, Taiwan (R.O.C.)
\and 
Taipei Astronomical Museum, 363 Jihe Rd., Shilin, Taipei 11160, Taiwan
\and 
Key Laboratory of Optical Astronomy, NAO, Chinese Academy of Sciences, 20A Datun Road, Beijing 100012, China
\and 
Astronomical Institute, Slovak Academy of Sciences, 059 60, Tatransk\'a Lomnica, Slovakia
\and 
Research School of Astronomy and Astrophysics, Australian National University, Canberra, ACT 2611, Australia
\and 
Main Astronomical Observatory of National Academy of Sciences of Ukraine, 27 Akademika Zabolotnoho St., 03680 Kyiv, Ukraine
\and 
Instituto de Astrof\'isica de Canarias, C/ v\'ia L\'actea, s/n, E-38205 La Laguna, Tenerife, Spain
\and 
Departamento de Astrof\'isica, Universidad de La Laguna, Av. Astrof\'isico Francisco S\'anchez, s/n, E-38206 La Laguna, Tenerife, Spain
\and 
Gettysburg College Observatory, Department of Physics, 300 North Washington St., Gettysburg, PA 17325, USA
\and 
Subaru Telescope, National Astronomical Observatory of Japan, 650 North A\`ohoku Place, Hilo, HI 96720, USA
\and 
Instituto de Astrofisica de Andalucia, CSIC, Apdo. 3004, 18080 Granada, Spain
\and 
Institute of Astronomy, University of Tokyo, 2-21-1 Osawa, Mitaka, Tokyo 181-0015, Japan
\and 
Dept. of Astronomy and Earth Science, Saitama University, 255 Shimo-Okubo, Sakura, Saitama 338-8570, Japan
\and 
Centro de Investigaciones de Astronomia, Apartado Postal 264, Merida 5101, Venezuela
\and
Cerro Tololo Inter-American Observatory (CTIO), Colina El Pino s/n, Casilla 603, La Serena, Chile
\and 
Astronomisches Institut, Ruhr-Universit\"at Bochum, Universit\"atsstr. 150, D-44801 Bochum, Germany
\and 
Instituto de Astronom\'{i}a, Universidad Cat\'{o}lica del Norte, Antofagasta, Chile
\and 
Physikalisches Institut, University of Bern, Sidlerstra\ss e 5, CH-3012 Bern, Switzerland
}

\received{2014 Feb 2}
\accepted{2014 Feb 21}
\publonline{2014 May 2}


\abstract{
We report our investigation of the first transiting planet candidate from the YETI project in the young ($\sim$4\,Myr old) open cluster Trumpler\,37. The transit-like signal detected in the lightcurve of the F8V star 2M21385603+5711345 repeats every $1.364894\pm0.000015$ days, and has a depth of $54.5\pm0.8$\,mmag in $R$. Membership to the cluster is supported by its mean radial velocity and location in the color-magnitude diagram, while the Li diagnostic and proper motion are inconclusive in this regard. Follow-up photometric monitoring and adaptive optics imaging allow us to rule out many possible blend scenarios, but our radial-velocity measurements show it to be an eclipsing single-lined spectroscopic binary with a late-type (mid-M) stellar companion, rather than one of planetary nature. The estimated mass of the companion is 0.15--0.44\,$M_{\sun}$. The search for planets around very young stars such as those targeted by the YETI survey remains of critical importance to understand the early stages of planet formation and evolution.
}

\maketitle



\section{Introduction}
\label{Sec:intro}

The transit technique has been one of the most successful methods for detecting extrasolar planets. It has enabled the discovery of more than 430\footnote{\url{http://exoplanet.eu} as of 2014-02-27} transiting exoplanets to date. The transit light curve enables one to determine the planet-to-star radius ratio and the inclination of the planetary orbit (Seager \& Mall{\'e}n-Ornelas \cite{sea03}), from which the absolute size of the planet can be established provided an estimate of the size of the parent star is available. Inferring the planetary mass usually requires high-precision radial velocity measurements that can be more expensive to obtain in terms of telescope time. These spectroscopic observations yield the minimum mass $M_p \sin i$ of the companion, which, combined with the inclination from the light curve and an estimate of the stellar mass, then enables the true mass $M_p$ of the planet to be calculated. Alternatively, measurements of transit timing variations can also yield the planetary mass in favorable cases.

The presence of transit-like signals in the light curve of a star is no guarantee of the planetary nature of the star's companion. The majority of such signals in ground-based transit searches end up being astrophysical false positives, and careful analysis is required to rule them out (see, e.g., Charbonneau et al.\ \cite{cha04}). For example, the transit of a small star in front of a much larger star (of earlier type, or a giant) can produce a transit depth indistinguishable from that of a true planet around a star. These scenarios can usually be discovered by performing low-resolution spectroscopy to classify the primary star. In the low-mass regime of degenerate compact objects the radius is independent of the mass (Guillot \cite{gui99}). Therefore, for a given transit signal, the companion may be a planet, a brown dwarf, or even a low-mass star. Medium- or high-resolution spectroscopy obtained near the quadratures is often sufficient to distinguish these cases, as companions that are brown dwarfs or late-type stars would generally induce easily detectable radial velocity variations on the star. The same spectra also permit the identification of cases of grazing eclipses of binary systems, which can also mimic the transit signals. Another important source of false positives are background eclipsing binaries blended with the foreground star (see, e.g., Torres et al.\ \cite{tor04}) . The light from the foreground target can reduce the otherwise deep eclipses of the binary to planetary proportions, mimicking a transit signal. In many cases adaptive optics imaging on large telescopes can help to reduce the possibility of such a contaminant at small angular separations.

Most transit surveys target field stars, for which age determination is generally challenging, and those searches are often biased towards main-sequence stars with ages of the order of Gyrs. No transiting planets have yet been identified around young stars (ages less than 100\,Myr), with the possible exception of the transit candidate around the weak-lined T~Tauri star CVSO30 in the 8\,Myr old cluster 25\,Ori (van Eyken et al.\ \cite{eyk12}, Barnes et al.\ \cite{bar13}). Only a few studies have searched for transiting planets around young stars. Two examples include the CoRoT satellite, which observed the 3\,Myr old cluster NGC\,2264 for 24 days (Affer et al.\ \cite{aff13}), and the MONITOR project, which is investigating several young clusters with ages in the range 1--200\,Myr (Hodgkin et al.\ \cite{hod06}), staring at each cluster for at least 10 nights. Neither project has reported any close-in planet discoveries to date.

This is rather surprising, as one expects planets to form already in the proto-planetary disk, which begins to dissipate a few Myr after star formation (Mamajek \cite{mam09}). Theoretical models are still rather uncertain at young ages as they depend upon the unknown initial conditions for planet formation (e.g., Marley et al.\ \cite{mar07}, Fortney et al.\ \cite{for08}, Spiegel \& Burrows \cite{spi12}). 
 Therefore, obtaining precise mass and radius measurements for planets around stars in young clusters of known age is critically important to test various aspects of current models of planet formation and evolution.

The YETI (Young Exoplanet Transit Initiative) network was established precisely to search for transiting planets in young clusters (Neuh\"auser et al.\ \cite{neu11}). The network consists of ground-based telescopes with apertures of 0.4\,m to 2\,m, spread out in longitude across several continents for significantly increased duty cycle and insurance against bad weather. The project is narrowly focused on clusters with ages of 2 to 20\,Myr, including Trumpler\,37, 25\,Ori, IC\,348, Collinder\,69, NGC\,1980, and NGC\,7243. Each of the clusters is monitored for three years at a time. In each year we schedule three YETI campaigns of one to two weeks duration each.

In this paper we report on the investigation of the first transiting planet candidate uncovered by the YETI project around the $R = 15$ magnitude star 2M21385603+5711345. This object is located in the area of the 4\,Myr old (Kun, Kiss \& Balog \cite{kkb08}) cluster Trumpler\,37, distant about 870\,pc from the Sun (Contreras et al.\ \cite{con02}). For a summary of the cluster properties we refer the reader to the work of Errmann et al.\ (\cite{err13}), which includes a list of candidates members of Trumpler\,37. The transit candidate studied here is not included in that list, so that its properties were not previously known. As we describe below, our extensive follow-up observations reveal it to be a false positive (an eclipsing single-lined spectroscopic binary with a low-mass stellar secondary) rather than a true planet. We use our observations to characterize both the companion and the parent star.


\section{Photometric detection}

\subsection{University Observatory Jena}
\label{Sec:gsh}

Monitoring of the Trumpler\,37 cluster began with the 90/60 cm Schmidt Telescope of the Astrophysical Institute and University Observatory Jena in July 2009, one year ahead of the first YETI campaign. The Schmidt Teleskop Kamera (STK, Mugrauer \& Berthold \cite{mug10}) was used in combination with a Bessell $R$ filter (Bessell \cite{bes90}). Exposure times were either 60\,s, 120\,s, or 105\,s. The increase from 60\,s to 120\,s yielded higher signal-to-noise ratios on fainter stars ($V\ge13$\,mag). After a recoating of the main mirror in October of 2011 the exposure time was set to 105\,s, on account of the improved reflectivity. We obtained data on Trumpler\,37 during 170 nights from the (northern) summer of 2009 until the fall of 2011.

The data were subjected to standard image reduction procedures, including overscan correction, dark current subtraction, and flat-fielding, using \textit{IRAF}\footnote{IRAF is distributed by the National Optical Astronomy Observatories, which are operated by the Association of Universities for Research in Astronomy, Inc., under cooperative agreement with the National Science Foundation.} routines.
A list of the pixel coordinates of each star was extracted using a mosaic of Trumpler\,37 and \textit{Source Extractor} (\textit{sextractor}; Bertin \& Arnouts \cite{ber96}), which is part of the \textit{GAIA}\footnote{GAIA is a derivative of the Skycat catalog and image display tool, developed as part of the VLT project at ESO. Skycat and GAIA are free software under the terms of the GNU copyright.} package. The final list contained 17097 stars and was used as the reference catalog for the photometry.

The tracking offsets were removed by extracting the star positions in each image with \textit{sextractor} and by comparing that list to the reference list of stars. The optimal size of the aperture used for the photometric measurements was determined separately for each night to account for changes in seeing and telescope focus. This was done by performing \textit{IRAF} photometry with 15 different aperture sizes in the range of 1 to 2.5 times the average FWHM (from \textit{sextractor}). To decrease computing time, we used only a subset of the 700 brightest stars that showed no variations in single nights, and only a subset of a maximum of 100 randomly derived images (including the first and last 10 images of each night). We calculated the standard deviation of the differential magnitudes between all stars for each of the 15 apertures, and selected the one giving the smallest sum of the standard deviations. Typical values for the aperture radii were between 3 and 4.5 pixels, corresponding to $4.5^{\prime\prime}$--$6^{\prime\prime}$. Aperture photometry was then performed with the optimized aperture, followed by differential photometry as described by Broeg, Fernandez \& Neuh\"auser (\cite{bro05}). All stars were used for the creation of an artificial standard star, but the more photometrically stable stars were weighted higher. The differential magnitudes $m_{\mathrm{diff}}$ were calculated in respect for this star. Since formal photometric errors are typically smaller than the dispersion of the light curves because they do not account for systematics, we have chosen to adjust the errors by rescaling them by a factor $f$ and adding a constant offset $s$ representing an error floor, to match the scatter of the light curves $\sigma_{\mathrm{lightcurve}}$:
$$s+f\cdot\Delta m_{\mathrm{inst},i,j}=\Delta m_{\mathrm{diff},i,j}\sim\sigma_{\mathrm{lightcurve},j}$$
with $i$, $j$ standing for image and star number (for further details see Broeg et al. (\cite{bro05})).
 The scale factor and floor were determined by testing a grid of values, in such a way as to obtain mean errors for the data points equal to the standard deviation of the light curve. To decrease computing time, only a subset of the brightest and most stable 400 stars was used for these calculations. Typical values for the scale and floor level were $f=0.95$ and $s=4.5$\,mmag (milli-magnitudes).

The images were reduced separately for each night. To combine data from different nights we corrected for differences in brightness between the standard stars with the following algorithm. For each star and night we calculated the median magnitude, and then the difference between these median magnitudes between two given nights. The median of all of these differences was then applied as an offset to the differential magnitudes of the second night.

\begin{figure}
 \centering
 \includegraphics[width=0.48\textwidth]{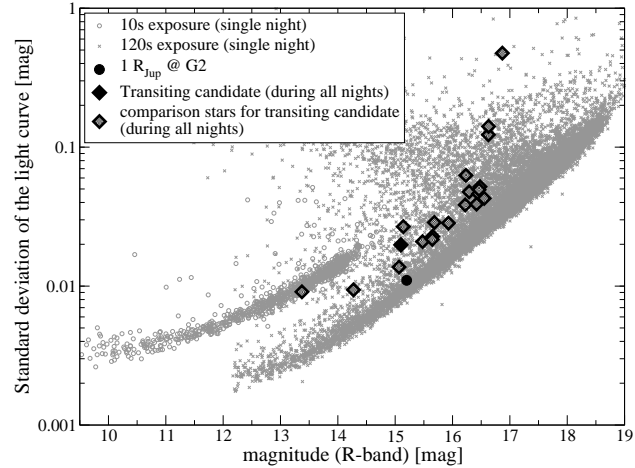}
  \caption{Photometric precision for a single night (2011 August 30) for all stars in the Jena field of view, as well as for all combined nights of the Jena data using only a subset of the comparison stars (see Sect.~\ref{Part:lightcurve-combi}). A total of 122 images were obtained at each exposure time on this night, and the precision reaches 2\,mmag for the brightest stars.}
 \label{Fig:phot_prec}
\end{figure}

The photometric precision of a light curve was calculated as the standard deviation of the brightness measurements. The results for a typical night are shown in Fig.~\ref{Fig:phot_prec}. For the brightest, non-saturated stars the photometric precision can be as low as 2\,mmag.

The apparent brightness of the stars in the field of view was determined by calibrating our differential magnitudes using stars in Trumpler\,37 from Sicilia-Aguilar et al.\ (\cite{sic04, sic05}). In order to obtain color information, observations on one night were carried out with Bessell $B$, $V$, $R$, and $I$ filters. 

\begin{figure}
 \includegraphics[width=0.48\textwidth]{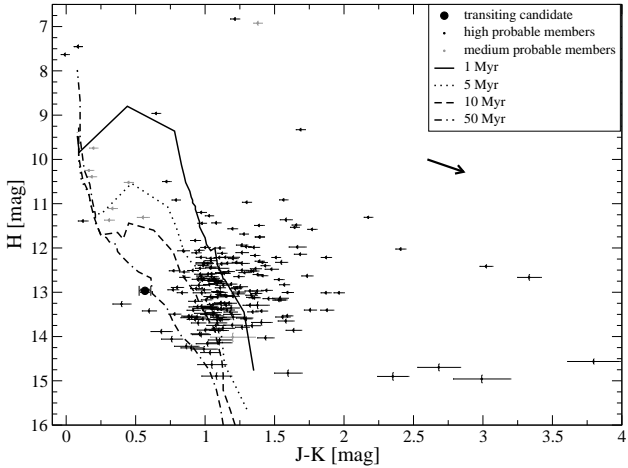}
 \caption{Color-magnitude diagram of Trumpler\,37 showing the location of the transit candidate (filled circle on the lower left). Other symbols represent stars with a high (black) or moderately high (gray) probability of membership following the work of Errmann et al.\ (\cite{err13}), ignoring stars with only proper motion membership. Also shown are solar metallicity model isochrones for 1, 5, 10, and 50\,Myr from Siess et al.\ (\cite{sie00}), adjusted to the adopted cluster distance of 870\,pc and average cluster visual extinction of 1.56\,mag. The arrow represents the reddening vector.}
 \label{Fig:cmd}
\end{figure}

The transit signal of the candidate was found by fitting a simple box model to the unfolded light curves, followed by subsequent visual inspection. The transit depth is $54.5\pm0.8$\,mmag, and the signals recur with a periodicity of about $1.365$ days. The parent star has a measured brightness of $R=15.08\pm0.12$\,mag. Its location in the color-magnitude diagram of the cluster is shown in Fig.~\ref{Fig:cmd}, along with that of other stars considered probable members of Trumpler\,37 (Errmann et al.\ \cite{err13}). The color and magnitude of the candidate place it near other cluster members in this diagram.
The folded and binned light curve of the star is shown in Fig.~\ref{Fig:lc_yeti}, together with data from other YETI telescopes (Sect.~\ref{Sec:YETI}).

\subsection{YETI network}
\label{Sec:YETI}

\begin{table}
 \centering
\caption{Dates of the YETI campaigns on Trumpler\,37. The last two rows give the number of participating telescopes, and the overall number of images gathered. Additional data were obtained by several telescopes outside of these runs, particularly the ones at Jena and Rozhen.}
\label{Tab:YETIruns}
\begin{tabular}{r r@{\hspace{1.5mm}}c@{\hspace{1.5mm}}c@{\hspace{1mm}}l r@{\hspace{1.5mm}}c@{\hspace{1.5mm}}c@{\hspace{1mm}}l}
\hline
           & \multicolumn{3}{r}{2010}& & \multicolumn{3}{r}{2011} &  \\
\hline
           & 3  & -- & 12 & Aug.        & 11 & -- & 22 & July \\
           & 26 Aug.& -- & 12 & Sept.       & 10 & -- & 22 & Aug. \\
           & 24 & -- & 30 & Sept.       & 9  & -- & 20 & Sept. \\
Telescopes &  \multicolumn{4}{c}{11}   & \multicolumn{4}{c}{9}   \\
Images     & \multicolumn{4}{c}{27500} & \multicolumn{4}{c}{21000}   \\
\hline
\end{tabular}
\end{table}

\begin{table*}
\begin{center}
\caption{List of the YETI telescopes and instruments that obtained data for this transit candidate. Sorting is from east to west starting in Asia. Telescopes for which the candidate was not in the field of view are not listed (Gunma/Japan, Byurakan/Armenia, Calar Alto/Spain, and Sierra Nevada/Spain).}
\label{Tab:YETI_teles}
\begin{tabular}{r lcl cc}
\hline
 Observatory     & Mirror dia- & CCD type        & no. of    & size of field          \\
                 & meter [m]   & (camera)        & pixels    & [$^\prime$ x $^\prime$]\\
\hline
 Lulin/Taiwan    & 1.00 (2)    & Marconi CCD36-40 PI1300B & $1340\times1300$ & $22\times22$     \\
                 & 0.41        & E2V 42-40 (U42)          & $2048\times2048$ & $28\times28$     \\
 Xinglong/China  & 0.90 (1)    & E2V CCD203-82            & $4096\times4096$ & $94\times94$     \\
 Rozhen/Bulgaria & 0.60 (2)    & FLI ProLine 09000        & $3056\times3056$ & $27\times27$     \\
                 & 0.70 (3)    & FLI ProLine 16803        & $4096\times4096$ & $73\times73$     \\
 Star\'a Lesn\'a/& 0.50        & SBIG ST10 MXE            & $2184\times1472$ & $20\times14$     \\
 Slovak Rep.     & 0.60        & SITe TK1024              & $1024\times1024$ & $11\times11$     \\
 Toru\'n/Poland  & 0.90 (1)    & SBIG STL-11000           & $4008\times2672$ & $48\times72$     \\
 Jena/Germany    & 0.90 (1)    & E2V CCD42-10 (STK)       & $2048\times2048$ & $53\times53$     \\
 Swarthmore/USA  & 0.62        & Apogee U16M KAF-16803    & $4096\times4096$ & $26\times26$     \\
 Gettysburg/USA  & 0.40        & SITe 003B                & $1024\times1024$ & $18\times18$     \\
 Tenagra II/USA  & 0.81        & SITe SI003 AP8p          & $1024\times1024$ & $15\times15$     \\
\hline                                                                                   
\end{tabular}
\end{center}

Notes: (1) 0.60\,m in Schmidt mode; 
(2) with focal reducer;
(3) 0.50\,m in Schmidt mode.
\end{table*}

\begin{figure}
 \includegraphics[width=0.48\textwidth]{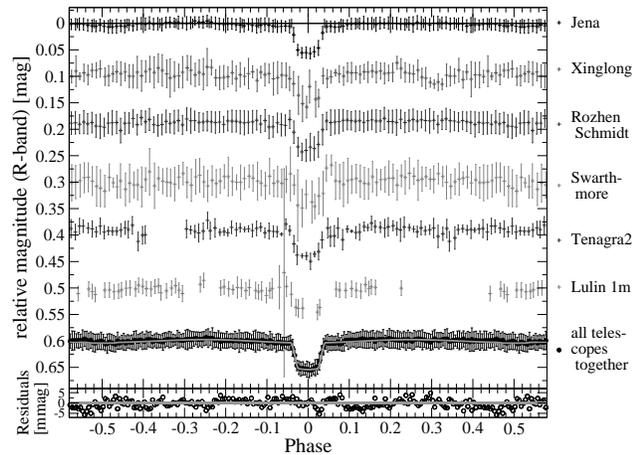}
 \caption{Phase-folded and binned light curve of the transit candidate in the $R$ band for individual telescopes, and for all telescopes combined. The ephemeris adopted is $P = 1.364894\pm0.000015$ days, $T_0 = 2455404.4932$ HJD (see Sect.~\ref{Sec:rv_lc_ana}). Bins are 0.01 in phase (0.005 for overall curve), and outliers were removed by sigma-clipping. For the light curves from individual telescopes the average number of points per bin is 124, 9, 54, 12, 4.6, and 1.6 (top to bottom). Jena has the smallest scatter (standard deviation $\sigma=3.6$\,mmag) due to the larger number of data points binned together. For the other telescopes $\sigma$ is 9.5, 4.9, 13.4, 7.6, and 7.0\,mmag. The combined light curve has an average of 100 data points per bin. The standard deviation is 3.1\,mmag (3.2\,mmag) for a bin size of 0.01 (0.005). The gray line through the combined light curve shows our best fit model described in Sect.~\ref{Sec:rv_lc_ana}.}
 \label{Fig:lc_yeti}
\end{figure}

The YETI observations were conducted during three campaigns each in the summer of 2010 and 2011 (see Table~\ref{Tab:YETIruns}).
The observations were carried out mostly in the $R$ band, although slight differences in the filters do exist among the different telescopes.
The data processing was similar to that described above for Jena, and included bias, dark, and flat-field correction, the determination of the optimal aperture, aperture photometry, and finally differential photometry. Te\-lescopes with smaller fields of view performed mosaicing. The phase-folded and binned light curves from the telescopes with the most data are shown in Fig.~\ref{Fig:lc_yeti}. Unfortunately, for some of the telescopes the transit candidate was located outside the field of view. The full list of participating sites that gathered data is given in Table~\ref{Tab:YETI_teles}. A map of all participating facilities in the YETI network may be seen in Fig.~1 of Errmann et al.\ (\cite{err14}). As shown in Fig.~3 of that work, the phase coverage enabled by this network is as high as 90\% for orbital periods up to 25 days.

\label{Part:lightcurve-combi}
In order to create a light curve combining all YETI data, we explored a number of procedures as follows: (1) Using the methodology for a single telescope as described in Sect.~\ref{Sec:gsh} resulted in unacceptable jumps of several tenths of a magnitude in the light curve, caused by slightly different filter transmission curves, color effects in the optics, and second-order airmass effects. (2) To avoid the latter two effects we used only a subsample of comparison stars with similar color and brightness as the target star, and that are close to it on the sky. (2a) Performing differential photometry night by night with this subsample and combining the results by recalculating the offsets as described in Sect.~\ref{Sec:gsh} still produced small offsets of a few percent, because the artificial standard stars were created using different weights for the stars in different nights. (2b) In another test we carried out differential photometry for this subsample of stars using the instrumental magnitudes from all nights and telescopes. This method resulted in the smallest deviation in the combined light curves between different telescopes, and is the one we finally adopted.
The criteria we used for selecting the subsample of comparison stars for our transit candidate are a maximum separation of 12\,arcmin, a brightness difference of no more than 1\,mag, and color differences less than 0.6\,mag in $B-V$, $V-R$ and $R-I$.

\section{Follow-up observations}

\subsection{Calar Alto 2.2~m telescope -- light curve}

\begin{figure}
 \includegraphics[width=0.48\textwidth]{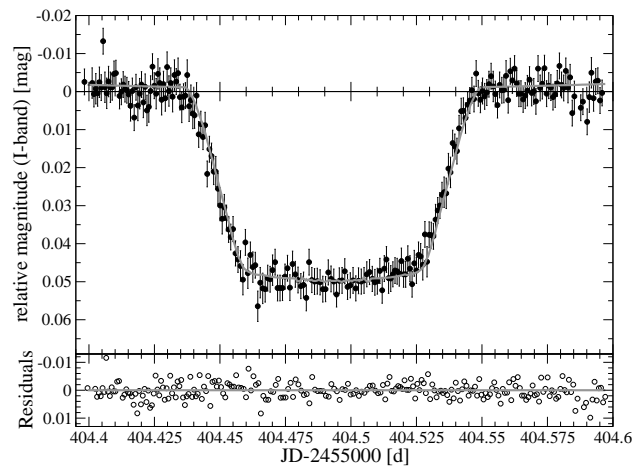}
 \caption{The $I$-band light curve of the transit candidate, taken with CAFOS on the Calar Alto 2.2\,m Telescope. The gray line drawn over the observations represents our best-fit model described in Sect.~\ref{Sec:rv_lc_ana}.}
 \label{Fig:lc_cafos}
\end{figure}

To improve upon the survey photometry from the YETI network, a high-precision light curve of the target star was obtained on 2010 July 26 (UT) with CAFOS (Calar Alto Faint Object Spectrograph) on the Calar Alto 2.2\,m telescope, using an $I$-band filter. Exposure times for imaging were 40\,s to 70\,s, depending on the atmospheric conditions. To decrease readout time we used $2\times2$ binning and recorded only a $201 \times 201$\,pixel sub-frame. We obtained a total of 238 images over 4.8 hours, including one hour of normal light before and after the transit. The images were bias-corrected, and flat-fielding was not necessary because of the stable auto-guiding (pixel shifts less than 0.5\,pixel), as shown by Maciejewski et al.\ (\cite{mac11}). Aperture photometry was performed on the 8 brightest non-saturated stars, including the target star. The optimal aperture was determined to be 2.44 pixels, or $2.59^{\prime\prime}$. The differential photometry (Broeg et al.\ \cite{bro05}) yielded standard deviations of 3.1 to 4.2\,mmag and similar weights (10-18\%) for the comparison stars. The final light curve is displayed in Fig.~\ref{Fig:lc_cafos}, and its shape is quite consistent with that of a typical transiting planet.

\subsection{Subaru telescope -- imaging}

High-resolution imaging can assist in ruling out eclipsing binaries blended with the target, which is particularly important for small telescopes with large PSFs (point-spread functions). To this end we obtained high-resolution images with the Infrared Camera and Spectrograph (IRCS, Koba\-yashi et al.\ \cite{kob00}) and adaptive optics AO188 at the Subaru telescope. Images in $J$, $H$, and $K$ were obtained with overall exposure times of 300\,s, 270\,s, and 250\,s, respectively.
 The observations were carried out on 2009 November 4 from 19:30 to 19:55 (local time) at airmass 1.265 to 1.278, and outside of the transit. The seeing was $0.4^{\prime\prime}$--$0.8^{\prime\prime}$. 

\begin{figure}
 \includegraphics[width=0.48\textwidth]{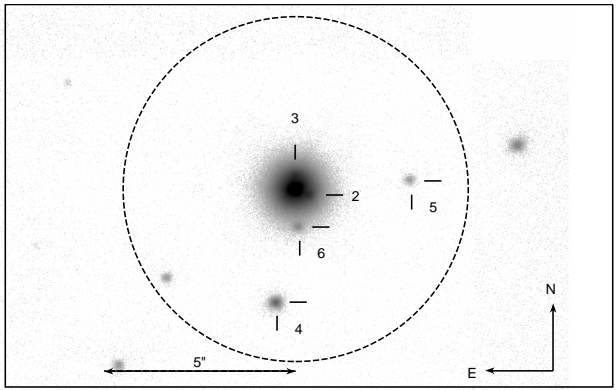}
  \caption{Reduced $H$-band image from IRCS at the Subaru telescope. Five faint sources are seen around the target star (identified as objects 2 to 5) that are unresolved in the YETI images. The dashed circle shows the typical YETI aperture for the photometry ($4.5^{\prime\prime}$). The faint source near the left edge of the circle was ignored, because for some observations it was outside of the aperture, and thus could not cause the transit on those nights.}
 \label{Fig:image_subaru}
\end{figure}

The master dark and flat field images were created with standard \textit{IRAF} routines. Further reduction was performed with the \textit{jitter} routine from the \textit{ESO eclipse software package} (Devillard \cite{dev97}), and consisted of dark and flat-field correction, sky background estimate and correction, and image combination. The final image for the $H$ band is shown in Fig.~\ref{Fig:image_subaru}. Several faint sources are detected within the typical $5^{\prime\prime}$ aperture used for the YETI photometry, and could in principle be causing the transit signal.

We performed PSF photometry to obtain an accurate measurement of the brightness of these close companions using \textit{StarFinder} (Diolaiti et al.\ \cite{dio00}). The best available PSF is that of the target star itself. To extract it, we first removed the neighboring sources as described next. We created an azimuthally averaged image of the target star, and subtracted it from the original. In this residual image (\textit{symrem}) we fitted the PSF of the well-resolved star~5 to the remaining stars, and produced a synthetic image containing only stars 2 to 6. Subtracting this synthetic image from the original one results in a less disturbed PSF of the main star (\textit{PSF\,1}). To improve upon this PSF we did the following: we fitted \textit{PSF\,1} to sources 2 to 6 in the \textit{symrem} image to create another image with an improved synthetic field. The new synthetic image was then subtracted from the original image with only the main star remaining. We used this to obtain a new \textit{PSF\,1}. This procedure was repeated until we saw no further improvements in the PSF of the target. The optimized PSF of the main star was then fitted to the original image to measure the instrumental brightness of the neighboring stars 2 to 6.

\begin{table*}
 \centering
\caption{Inferred brightness of the sources in Fig.~\ref{Fig:image_subaru}. The first line labeled ``2MASS'' gives the values for the combined light of all sources in the aperture as listed in the 2MASS catalog (converted to the instrumental passbands). Individual magnitudes for sources 2--6 were derived as described in the text. Uncertainties are given in parentheses in units of the last significant digits. Star 3 was not resolved in the $J$ band, due to less efficient AO. 
The $R$ magnitudes were extrapolated from the near-infrared colors, using the intrinsic colors as listed by Kenyon \& Hartmann (\cite{ken95}) and the interstellar extinction law of Rieke \& Lebofsky (\cite{rie85}). Depending on the (main-sequence) spectral type, each source has a range of possible $R$ magnitudes. The uncertainties in the $R$ band can be as large as 1.1\,mag.}
\label{Tab:ao_bright}
\begin{tabular}{l ccc ll}\hline
      & $J$ [mag]   & $H$ [mag]   & $K$ [mag]   &  $R$ [mag]    & angular sep.  \\
      &  \multicolumn{3}{c}{measured}           &  extrapolated & [$^{\prime\prime}$]       \\     
\hline
2MASS & 13.454 (29) & 12.968 (35) & 12.925 (40) &               &         \\
\hline
1     & 13.503 (29) & 13.053 (35) & 12.925 (40) &               &         \\
2     & 17.761 (34) & 16.881 (36) & 16.427 (41) & 20.7--22.9    & 0.43   \\
3     &             & 17.411 (37) & 17.187 (43) & 19.4--21.1    & 0.30   \\
4     & 17.838 (30) & 17.245 (36) & 17.016 (41) & 19.4--21.6    & 3.02    \\
5     & 19.231 (37) & 18.552 (40) & 18.105 (48) & 21.5--23.8    & 3.00    \\
6     & 19.651 (57) & 18.613 (42) & 18.224 (52) & 22.9--25.1    & 0.98   \\
\hline
\end{tabular}
\end{table*}

Apparent magnitudes for all sources including the target were computed from the measured magnitude differences with the constraint that the combined brightness must match the magnitudes listed in the 2MASS catalog (Skrutskie et al.\ \cite{skr06}). The 2MASS $K_s$ magnitudes were converted to the instrumental $K$-band system used with the IRCS following Carpenter (\cite{car03}). Table~\ref{Tab:ao_bright} presents our derived $J$, $H$, and $K$ magnitudes. However, the more relevant passband for our purposes is the $R$ band. Because the spectral type, extinction, and distance to each of the faint sources is unknown, we estimated $R$-band magnitudes for them by exploring a range of possible spectral types using our measured near-infrared colors as a guide, along with the intrinsic colors of main-sequence stars as listed by Kenyon \& Hartmann (\cite{ken95}). Extinction and distance were calculated for each spectral type using the interstellar extinction law of Rieke \& Lebofsky (\cite{rie85}). We used the Kenyon \& Hartmann table to infer the brightness for each spectral type in the $R$ band. From these experiments the brightest that any of these close companions can be in that passband is $19.4\pm1.1$\,mag. In the worst case scenario, if this source is a fully eclipsing binary composed of two identical main-sequence stars (maximizing the eclipse depth), the depth of the transit-like signal would be $4.3^{+7.4}_{-2.7}$\,mmag. Since this is much smaller than the measured depth of $\sim$50\,mmag, we can rule out all of these sources as possible contaminants. In order to mimic the observed transit, the binary would have to be as bright as $R = 16.5$\,mag. Other companions that are angularly much closer to the target and are not seen in Fig.~\ref{Fig:image_subaru} may still be present, of course, but we have no means of ruling them out.

\subsection{Spectroscopy}

\subsubsection{Calar Alto 2.2~m -- low-resolution spectrum}

\begin{figure*}
 \centering
 \includegraphics[width=0.98\textwidth]{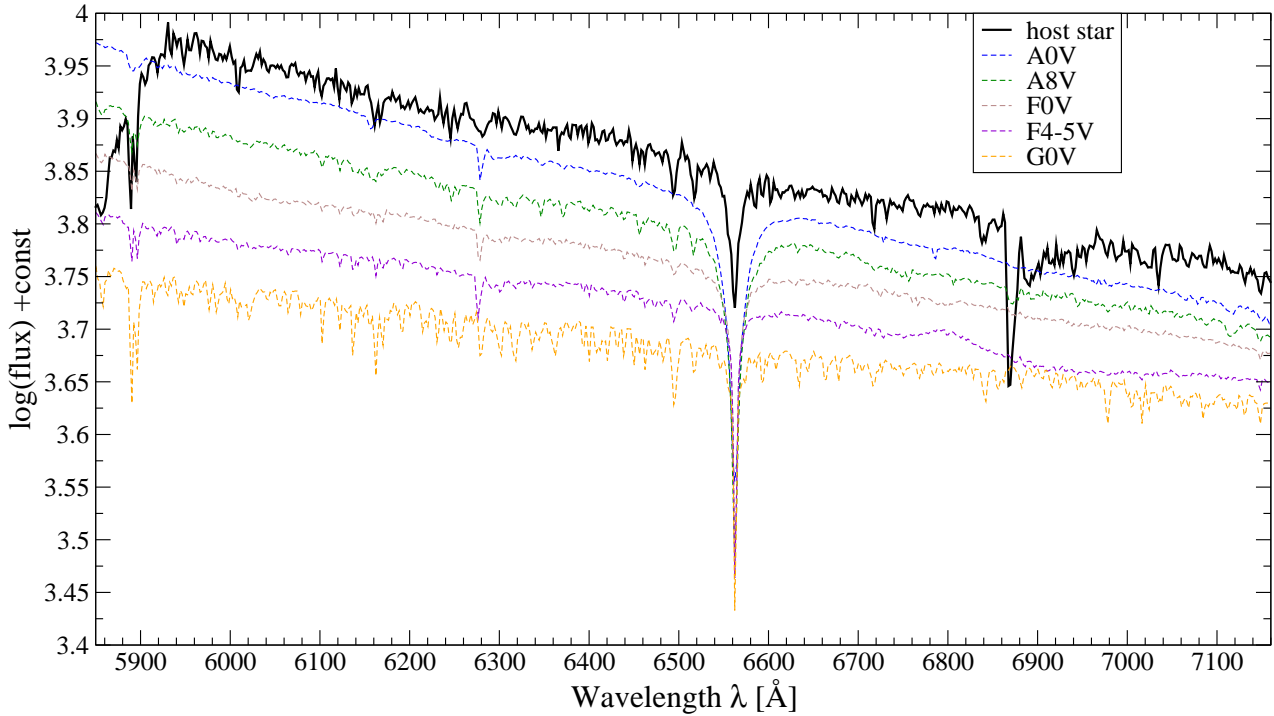}
 \caption{Spectrum of the transit candidate host star taken with CAFOS at the Calar Alto 2.2\,m Telescope, compared against standard stars over a range of spectral types (thin dashed lines). The black line represents the spectrum of the target corrected for atmospheric extinction using the standard star HD\,217086 (see text). The sharp drop at the blue edge is caused by reduced flux on the detector, which was not corrected with the standards. Telluric absorption is visible in the wavelength range 6865--6970\,\AA.}
 \label{Fig:spec_cafos}
\end{figure*}

Two low-resolution spectra ($R=1800$) were obtained with CAFOS on the 2.2\,m Calar Alto telescope on 2010 August 26 (UT) to aid in the spectral classification of the target. The spectral coverage is $\sim$5800--7200\,\AA, and the exposure time was 1800\,s for both observations. Bias correction was applied, and the spectra were extracted with the \textit{IRAF} task \textit{apall}. Wavelength calibration was based on a Hg-He-Rb arc exposure from the same night, and cosmic rays were removed by hand. The two spectra were then combined. Additionally, we obtained a spectrum of the O7V star HD\,217086 at the same airmass for atmospheric correction; the difference in airmass was smaller than 0.04. The differences between the continuum slope of the spectrum of HD\,217086 and that of published spectra of the standard O7V star HD\,47839 (Valdes et al.\ \cite{val04}) and of the O7.5V star HD\,44811 (Silva \& Cornell \cite{sil92}) were applied to the spectrum of the target star to remove the effects of atmospheric extinction. There is no significant difference between the results using both literature standards.
The corrected spectrum is shown in Fig.~\ref{Fig:spec_cafos}; further discussion follows in the analysis section below (Sect.~\ref{Sec:spectraltype}). 

\subsubsection{Keck telescope -- high-resolution spectra}

\begin{figure}
 \includegraphics[width=0.48\textwidth]{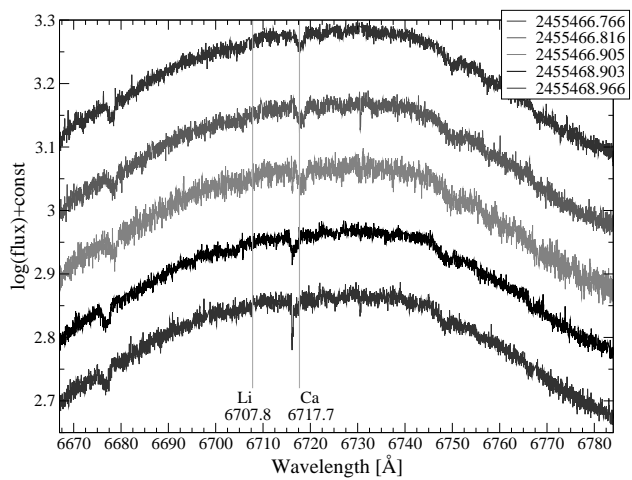}
  \caption{Keck spectra of our target in the Ca~6717.7\,\AA\ region, offset vertically for clarity. The Doppler shift of this line from night to night is clearly visible. The location of the Li~6707.8\,\AA\ line is also indicated, but this feature is not detected in our spectra.}
 \label{Fig:Keck-spec}
\end{figure}

The target star 2M21385603+5711345 was also observed spectroscopically with the High Resolution Echelle Spectrometer (HIRES, Vogt et al.\ \cite{vog94}) at the Keck\,I telescope, for the purpose of monitoring its radial velocity (RV). The resolving power of the instrument with the setup we use is approximately $R = 55,000$, and the spectral coverage is $\sim$3800--8000\,\AA, rec\-ord\-ed on three contiguous CCDs. Five spectra were obtained on 2010 September 26 and September 28, close to the expected RV extrema based on the ephemer\-is. The exposure time was 4200\,s for each spectrum. Exposures of a Th-Ar lamp were taken before and after the science exposures for the purpose of wavelength calibration. Data reduction including dark, flat-field, and cosmic-ray correction was carried out using standard procedures, followed by extraction of the spectra.
 Fig.~\ref{Fig:Keck-spec} shows one of the orders in each our spectra containing the Ca~6717.7\AA\ line. No sign of another star is seen in these spectra.

Radial velocities from the Keck spectra were derived by cross-correlation using the \textit{IRAF} task \textit{xcsao} (Kurtz \& Mink \cite{kur98}). We used a synthetic template based on Kurucz model atmospheres with parameters approximating those of the target star: $T_{\mathrm{eff}} = 6000$\,K, $\log g=4.5$, $v\sin i=30$\,km\,s$^{-1}$, and $\mathrm{[Fe/H]}=0$ (see Sect.~\ref{Sec:spectraltype}).
The RV at each epoch was derived as a weighted mean of 12 independent radial velocity measurements from the usable echelle orders on the red CCD. We rejected orders with very few stellar lines or that were severely contaminated by telluric lines. The wavelength range of the red chip was not covered by the model spectrum. The RV uncertainty for each epoch was taken to be the standard deviation of the 12 measurements. Errors from cross correlation in an individual order are about 2\,km\,s$^{-1}$; the standard deviations of all 12 orders ranged from 1.3 to 1.8\,km\,s$^{-1}$. 
Barycentric corrections were applied for the flux-weighted mid-time of each exposure, as provided by the exposure meter of the instrument (Kibrick et al.\ \cite{kib06}). The radial velocity standard star HD\,182488 (adopted ${\rm RV} = -21.508$\,km\,s$^{-1}$) was observed on each night to correct for instrumental drifts. The resulting RVs are given in Table~\ref{Tab:RV_Keck}. The large amplitude of the velocity variations clearly indicates that the companion is of stellar mass, thus ruling out the planetary nature of the candidate. Below we provide evidence that the companion is a late-type star.

\begin{table}
 \centering
\caption{Barycentric radial velocities of the target from HIRES.}
\label{Tab:RV_Keck}
\begin{tabular}{ccc}
\hline
HJD-2455000 & RV [km\,s$^{-1}$] & standard deviation [km\,s$^{-1}$]\\
\hline
466.7709   &  11.5      & 1.6 \\
466.8208   &  16.6      & 1.3 \\
466.9105   &  21.5      & 1.8 \\
468.9080   &  $-49.0$   & 1.4 \\
468.9715   &  $-51.2$   & 1.5 \\
\hline
\end{tabular}
\end{table}

\subsubsection{MMT -- high-resolution spectrum}	
\begin{sloppypar} \tolerance 9999
An additional spectrum of the star was re\-cord\-ed with the Hectochelle instrument (Szentgyorgyi et al.\ \cite{sze98}) on the MMT on 2010 November 21 (UT). Hectochelle is a multi-object fiber-fed echelle spectrograph with a spectral range of 155\,\AA\ centered at a wavelength of 5225\,\AA\ (setup RV31), which includes the Mg\,b triplet. The exposure time was 1800\,s. Extraction was performed using the custom pipeline written at the Harvard-Smithsonian Center for Astrophysics\footnote{\url{http://tdc-www.harvard.edu/instruments/hectochelle/pipeline.html}}. Unfortunately no RV standard was observed on the same night, which prevents us from deriving an accurate radial velocity from this observation. Nevertheless, the spectrum is useful for inferring the stellar properties, as described below in Sect.~\ref{Sec:spectraltype}.
\end{sloppypar}

\section{Properties of the host star}

Because both the light curves and the radial velocities only allow one to obtain the properties of the companion relative to those of the primary star, we describe in this section our efforts to characterize the star with the observations we have in hand. Table~\ref{Tab:host_star} summarizes the observed properties, as well as the derived properties discussed below.

\begin{table}
 \centering
\caption{Properties of the host star.}
\label{Tab:host_star}
\begin{tabular}{l c  l}
\hline
Property & Value    & Com.  \\
\hline
                           & 2M21385603+5711345   &          \\
$\alpha$ (J2000) [h:m:s]   & 21:38:56.1           &          \\
$\delta$ (J2000) [$^\circ$:$^\prime$:$^{\prime\prime}$] & +57:11:34 &  \\
 $B$ [mag]                 & $16.640 \pm 0.056$ & [1] \\
 $V$ [mag]                 & $15.569 \pm 0.046$ & [1]      \\
 $R$ [mag]                 & $15.08 \pm 0.12$ & [1]      \\
 $I$ [mag]                 & $14.47  \pm 0.10$  & [1] \\
2MASS $J$ [mag]            & $13.454\pm0.029$ & [2]      \\
2MASS $H$ [mag]            & $12.968\pm0.035$ & [2]      \\
2MASS $K_s$ [mag] & $12.887\pm0.030$ & [2]      \\
EW(Li~I) [m\AA]           & $<5$             & [1]      \\
Spec.\ Type                 & F8               & [1]      \\
$T_\mathrm{eff}$ [K]       & $6120\pm450$     & [1]      \\
\hline
\end{tabular}

Notes: 
[1]: this work, 
[2]: Skrutskie et al.\ \cite{skr06}. 
\end{table}

\subsection{Spectral type and rotation}
\label{Sec:spectraltype}

An initial estimate of the spectral type of the star was obtained by comparing its colors with tables of standard main-sequence colors from the literature. We assume hereafter that the stellar companion inferred above from the RV variations is faint enough that it does not significantly affect the measured magnitudes. For this exercise we used the brightness measurements from the 2MASS catalog and our own $BVRI$ measurements. Corrections for extinction were applied based on the mean visual extinction to the cluster of $A_V = 1.56 \pm 0.55$\,mag (Mercer et al.\ \cite{mer09}). The intrinsic colors in Table~5 of Kenyon \& Hartmann (\cite{ken95}) suggest a spectral type around F7 or F9. For a more quantitative estimate we computed six non-independent color indices from the photometry, and applied appropriate reddening corrections following Cardelli et al.\ (\cite{Car89}). We used the color-temperature calibrations of Casagrande et al.\ (\cite{cas10}), based on the infrared flux method, to obtain effective temperatures from each index. The weighted average temperature is $6100 \pm 250$\,K, consistent with the above classification.

A rough estimate of the spectral type was attempted also using the CAFOS spectrum, with mixed results. Comparing the slope of the continuum with that of standard stars from Valdes et al.\ (\cite{val04}) suggests a much earlier spectral type than inferred above (see Fig.~\ref{Fig:spec_cafos}), while the measurement of the equivalent widths of key features such as H$\alpha$, the Na~I D lines (5890\,\AA, 5896\,\AA), and the Fe line at 6495\,\AA\ are more consistent with an F or G star. A similar classification was obtained by measuring other lines including the Ca~I line at 6162\,\AA\ with \textit{SPTCLASS} (\textit{SPecTral CLASSificator code}\footnote{\url{http://dept.astro.lsa.umich.edu/\~hernandj/SPTclass/sptclass.html}}, Hernandez et al.\ \cite{herna}).

Another, more reliable spectroscopic estimate was obtained from the Hectochelle spectrum, by cross-correlating it against a grid of synthetic spectra based on model atmospheres by R.\ L.\ Kurucz. Cross-correlations were performed with the \textit{IRAF} task \textit{xcsao}, and the calculated spectra spanned a broad range of effective temperatures, surface gravities, and rotational velocities, for an assumed solar metallicity. Peak cross-corr\-elation coefficients reached values as high as 92--93\%, and a total of 168 synthetic spectra were found to provide equally satisfactory fits. The average and standard deviation of the parameters of these 168 spectra are $T_{\mathrm{eff}}=6120\pm450$\,K, $\log g=4.28\pm0.53$, and $v\sin i=44\pm11$\,km\,s$^{-1}$. We adopt these values for the remainder of the analysis. The temperature corresponds to a spectral type of F8V, in good agreement with most of our other estimates. The uncertainty yields a possible range from F4 to G6.

These properties correspond broadly to a main-sequence star with a mass in the range 0.90--1.33\,$M_{\sun}$ and a radius of 0.91--1.35\,$R_{\sun}$. Interestingly, we find that the measured projected rotational velocity of 44\,km\,s$^{-1}$ would be consistent with synchronous rotation in a binary with the observed orbital period of 1.365\,days if the primary had a radius of 1.2\,$R_{\sun}$. This happens to be near the middle of the radius range given above, and is another indication that the companion is unlikely to be of planetary nature, because its mass would be too small to induce the tidal forces needed to lock the primary into synchronous rotation.

\subsection{Lithium}

No evidence for the lithium 6707.8\,\AA\ line is seen in either our low-resolution or high-resolution spectra, with an upper limit on its equivalent width of 5\,m\AA. This is smaller than expected for a member of Trumpler\,37, and would cast doubt on the membership of the star. Indeed, among the probable cluster members listed by Errmann et al.\ (\cite{err13}), the star with the earliest spectral type with measured lithium absorption is F9, and its strength is $\mathrm{EW(Li)} = 100$\,m\AA. A few early G-type stars reach EWs up to 600\,m\AA. 
The stellar (as opposed to planetary) nature of the binary companion indicated earlier by our RV measurements suggests that the absence of the Li line could perhaps be due to stronger mixing in the primary star caused by the secondary, resulting in faster depletion of Li. Increased Li burning could also have occurred as a result of episodic accretion (Baraffe \& Chabrier \cite{bar10}). In any case, the late-type companion itself would be expected to have strong lithium ($\mathrm{EW(Li)}\sim900$\,m\AA), as other M stars in the cluster do. However, dilution by the light of the much brighter primary star would render the feature undetectable in our spectra. In conclusion, the Li diagnostic in our target is inconclusive regarding membership in the Trumpler\,37 cluster.

\subsection{Proper motion}

Finally, in Table~\ref{Tab:pm} we list the proper motion (p.m.) values according to the UCAC4 (Zacharias et al.\ \cite{zac13}) and PPMXL catalogs (Roeser, Demleitner \& Schilbach \cite{roe10}). For easier comparison we have also transformed these values to the reference frame for the cluster proper motions adopted by Marschall \& van Altena (\cite{mar87}), which is aligned with the minor and major axes of the Trumpler\,37 p.m.\ distribution (a rotation by $48^\circ$). Though formally consistent with the cluster motion, the uncertainties are large enough that our target is equally likely to be a cluster member or a background star.

\begin{table}
 \centering
\caption{Proper motion of our target compared to the cluster p.m.\ from Marschall \& van Altena (\cite{mar87}). Components $\mu_X$ and $\mu_Y$ give the values rotated by $48^\circ$ to transform them to the reference frame used by those authors, in which the axes are parallel to the minor and major axes of the cluster p.m.\ distribution. All values are given in units of mas/yr.}
\label{Tab:pm}
{\footnotesize
\begin{tabular}{r c@{\hspace{1.5mm}}c c@{\hspace{1.5mm}}c}
\hline
          & $\mu_\alpha$ & $\mu_\delta$  & $\mu_X$       & $\mu_Y$   \\
\hline
 UCAC4    & $+0.1\pm3.4$ & $-2.4\pm3.5$  & $+1.9\pm4.9$  & $-1.5\pm4.9$  \\
 PPMXL    & $-4.4\pm4.0$ & $+1.9\pm4.0$  & $-4.4\pm5.6$  & $-2.0\pm5.6$  \\
\hline
 Cluster  &              &               & $+0.1\pm1.9$  & $-0.2\pm2.8$   \\
\hline
\end{tabular}
}
\end{table}

\section{Radial velocity and light curve analysis}
\label{Sec:rv_lc_ana}

A refined orbital period for the system was derived using the string-length algorithm of Dworetsky (\cite{dwo83}), as implemented in the program \textit{stringlength} by Broeg (\cite{bro06}). We obtained $P=1.364894 \pm 0.000015$\,d, in which the uncertainty corresponds to the range over which additional significant signals are seen.  We created a high quality light curve for initial analysis by folding the Jena photometry with the period $P$ and applying a binning of $0.002$ in phase. The depth and duration of the transit as measured using the software available with the \textit{Exoplanet Transit Database (ETD)} (Poddan{\'y}, Br{\'a}t \& Pejcha \cite{pod10}) are listed in Table~\ref{Tab:params_lightcurve}, both for the Jena data and the Calar Alto $I$-band photometry.

\begin{table}
 \centering
\caption{Initial light-curve parameters from ETD and TAP.}
\label{Tab:params_lightcurve}
\begin{tabular}{r cc}
\hline
                              & Jena ($R$)           & Calar Alto ($I$)   \\
                              & \multicolumn{2}{c}{\textit{ETD}} \\
\hline
Depth                  [mmag] & $54.5\pm0.8$       & $51.8\pm0.5$  \\
Duration $t_{\mathrm{trans}}$ [min] & $162.7\pm1.5$       & $159.8\pm0.8$  \\

                              &  \multicolumn{2}{c}{\textit{TAP}}   \\
\hline
$i$ [$^\circ$]                & $85.1^{+2.5}_{-1.9}$         & $85.2^{+2.5}_{-1.9}$\\
$a/R_\mathrm{A}$              & $4.35^{+0.23}_{-0.24}$       & $4.36^{+0.22}_{-0.24}$\\
$R_\mathrm{B}$/$R_\mathrm{A}$ & $0.2023^{+0.0056}_{-0.0059}$ & $0.2019 ^{+0.0058}_{-0.0060}$\\
\hline
\end{tabular}
Note: Durations are measured between the first and last contacts.
\end{table}

Further analysis of the light curves was performed with the \textit{Transit Analysis Package (TAP)}, version 2.104 (Gazak et al.\ \cite{gaz12}), which we used to determine the inclination angle $i$, the relative semi-major axis in units of the stellar radius $a/R_\mathrm{A}$, and the radius ratio between the companion (B) and the main star (A), $R_\mathrm{B}$/$R_\mathrm{A}$. The period was held fixed at the value reported above, and the quadratic limb-darkening coefficients were allowed to vary freely. A circular orbit was assumed. The resulting parameters for the Jena $R$-band data and the Calar Alto $I$-band data are consistent (see Table~\ref{Tab:params_lightcurve}).

Even though it is eclipsing, the fact that our target is only a single-lined spectroscopic binary implies that the absolute masses of the stars cannot be obtained directly from the data, particularly that of the secondary, which would speak to its nature. Nevertheless, a weak constraint is contained in the photometry, principally through the out-of-eclipse variations, which depend on the mass ratio and are caused in part by tidal deformation of the stars. We explored this weak dependence by fitting all of the photometry (from both YETI and Calar Alto) and the radial velocities simultaneously with the \textit{PHOEBE}\footnote{\url{http://phoebe-project.org/}} program (PHysics Of Eclipsing BinariEs, version 0.31a, Pr{\v s}a \& Zwitter \cite{prv05}), to obtain the plausible range of mass ratios allowed by the data. We adopted a model for a detached eclipsing binary, with albedo, limb-darkening (for a linear law), and gravity-brightening parameters appropriate for stars with convective envelopes. Initial values for the various fitting parameters were derived from the \textit{TAP} solutions above, along with the period reported earlier. The temperature of the primary star was set to 6100\,K. This exercise allowed us to restrict the mass ratio to the range between 0.17 and 0.29, which, together with the range of primary masses established in Sect.~\ref{Sec:spectraltype}, yields a mass for the companion of $M_\mathrm{B} = 0.15$--0.44\,$M_{\sun}$. Plausible radius ratios were found to be in the range $R_{\rm B}/R_{\rm A} = 0.205$--0.209, slightly larger than determined with the less sophisticated \textit{TAP} procedure. The corresponding radius interval for the low-mass companion is $R_\mathrm{B} = 0.18$--0.30\,$R_{\sun}$. Assuming the secondary is on the main sequence, these values are typical of a mid-M star. The range of effective temperatures derived for the secondary is consistent with this classification. We report these and other results in Table~\ref{Tab:params_phoebe}. 

\begin{figure}
 \includegraphics[width=0.48\textwidth]{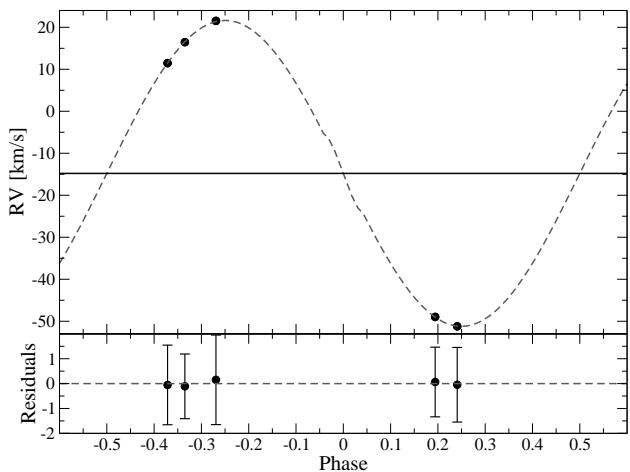}
 \caption{Keck radial velocity measurements of our target, along with the best-fit circular model. The velocity semi-amplitude is $36.5$\,km\,s$^{-1}$, and the horizontal line represents the center-of-mass velocity $\gamma = -14.79$\,km\,s$^{-1}$. The standard deviation of the fit is $0.32$\,km\,s$^{-1}$. Residuals are shown at the bottom.}
 \label{Fig:rv_orbit}
\end{figure}

A graphical representation of our best-fit model for the light curve may be seen overplotted on the photometry of Fig.~\ref{Fig:lc_yeti} and Fig.~\ref{Fig:lc_cafos}, and the fit to the radial velocities is illustrated in Fig.~\ref{Fig:rv_orbit}. Small out-of-eclipse changes caused by the tidal bulge of the primary are visible. The center-of-mass velocity we infer from the measurements is $\gamma = -14.79 \pm 0.09$\,km\,s$^{-1}$. This is in good agreement with the mean radial velocity reported for Trumpler\,37 of $-15.3\pm3.6$\,km\,s$^{-1}$ (Sicilia-Aguilar et al.\ \cite{sic06-2}), which supports membership in the cluster.

\begin{table}
\caption{Results of our joint photometric-spectroscopic fits with \textit{PHOEBE}.}
\label{Tab:params_phoebe}
\begin{center}
\begin{tabular}{r c}
\hline
                &  \textit{PHOEBE}                \\
\hline
$R_\mathrm{B}$/$R_\mathrm{A}$(*) & 0.205--0.209        \\
$M_\mathrm{B}$/$M_\mathrm{A}$    & 0.171--0.288        \\
$a/R_{\sun}$                     & 4.42--6.74         \\
$T_{\mathrm{eff},2}$ [K]         & 2827--3549         \\
$\log(g)_1$                      & 4.14--4.36         \\
$\log(g)_2$                      & 4.95--4.98         \\
$T_0$ [HJD]            &   $2455404.4932\pm0.0002$     \\
$\gamma$ [km\,s$^{-1}$] &   $-14.79\pm0.09$              \\
\hline
\end{tabular}
\end{center}
Notes: ``A'' denotes the brighter and more massive star, ``B'' the companion, $\gamma$ is the center-of-mass velocity. (*) indicates values which were calculated from other results of \textit{PHOEBE}.
\end{table}

The above analysis assumed no light contribution from the other stars visible in photometric aperture (see Fig.~\ref{Fig:image_subaru}). This could in principle affect the results, leading, e.g., to an overestimate of the transit depth and consequently of the radius ratio. While the spectral types of these nearby stars are not known, we estimate their contribution to the overall flux in the passbands of interest to be less than 2.2\%. This level of contamination has only a minimal impact on the parameters given in Table~\ref{Tab:params_phoebe}.

\section{Conclusions}

We have performed extensive follow-up observations (photometric monitoring, imaging, and spectroscopy) of the first transiting planet candidate (2M21385603+5711345) from the YETI network, in the young open cluster Trumpler\,37. Membership in the cluster seems likely based on its mean radial velocity and location in the color-magnitude diagram, though other evidence (small proper motion and lack of Li $\lambda$6708 absorption) is inconclusive.

Careful analysis of our survey and follow-up observations shows that the candidate is an astrophysical false positive rather than a true planet. We determine the companion to be a late-type (mid-M) star in an eclipsing configuration around the late F primary star. Close visual inspection of the fully reduced and rebinned YETI light curve in Fig.~\ref{Fig:lc_yeti} after combining all telescopes shows a hint of a secondary eclipse at phase 0.5 that would normally be an early warning sign, but that is too subtle to have been noticed earlier in the analysis.

\begin{sloppypar}
\tolerance 9999
While disappointing, this outcome is not surprising given the fact that all other ground-based transit surveys have experienced very high rates of false positives typically in excess of 80\% (see, e.g., Brown \cite{Brown03}, Konacki et al.\ \cite{Kon03}, O'Donovan et al.\ \cite{Odon06}, Latham et al.\ \cite{Lat09}). Regardless of this result, the search for planets around very young stars remains of critical importance for our understanding of the formation and evolution of exoplanets, and to learn about the properties of these objects at the very early stages.
\end{sloppypar}

With the YETI network in full operation we continue to monitor several young open clusters as described in Sect.~\ref{Sec:intro}, and follow-up observations for two additional transit candidates are currently underway.

\acknowledgements
All participating observatories appreciate the logistic and financial support of their institutions and in particular their technical workshops.

We wish to thank the following persons for participating in the observations of Trumpler\,37: E.L.N.\ Jensen and D.H.\ Cohen at the Swarthmore telescope; A.\ Niedzielski at the Torun telescope; E.H.\ Nikogossian, T.\ Movsessian, and H.\ Harutyunyan at the Byurakan telescope; and T.\ Roell, S. Baar, E.\ Schmidt, F.\ Gie{\ss}ler, H.\ Gilbert, I.\ H\"ausler, D.\ Keeley, and P.\ N\"ahrlich for the observations at the Jena telescope.
We are grateful to D.W.\ Latham for the helpful discussion and proposing observations.

RN, MK and RE would like to thank DFG for support in the Priority Programme SPP 1385 on the {\em First ten Million years of the Solar System} in project NE 515 / 34-1 and 34-2. 
BD, JGS, MMH, LT, TE, and RN thank DFG in SFB TR 7 in TPs C2, C7, and B9. 
CM and KS thank DFG in project SCHR 665 / 7-1. 
CG and MM thank DFG in project MU 2695 / 13-1. 
CA thanks DFG in project NE 515 / 35-1 and 35-2 in SPP 1385. 
SR and TR thanks DFG in project NE 515 / 33-1 and 33-2 in SPP 1385. 
NP thanks DFG in project KR 2164 / 10-1. 
DD acknowledges the partial financial support of the project DDVU 02/40-2010 of the
Bulgarian National Science Fund. 
Wu, Z.Y.\ was supported by the Chinese National Natural Science Foundation grant No.\ 11373033. This work was also supported by the joint fund of Astronomy of the National Nature Science Foundation of China and the Chinese Academy of Science, under Grant U1231113. 
Zhou, X. was supported by the Chinese National Natural Science Foundation grands No. 11073032, and by the National Basic Research Program of China (973 Program), No. 2014CB845704 and 2013CB834902. 
This work has been supported by a VEGA Grant 2/0143/13 of the Slovak Academy of Sciences. 
We would like to acknowledge financial support from the Thuringian government (B 515-07010) for the STK CCD camera used in this project. 
The authors wish to recognize and acknowledge the very significant cultural role and reverence that the summit of Mauna Kea has always had within the indigenous Hawaiian community.  We are most fortunate to have the opportunity to conduct observations from this mountain.


\end{document}